\documentclass[preprint,12pt]{elsarticle}
\usepackage{amssymb}
\usepackage{graphicx}

\journal{Elsevier}

\begin{document}

\title{ Theoretical investigations of electronic structure and magnetism in   Zr\raisebox{-.2ex}{\scriptsize 2}CoSn  full-Heusler compound}
\author[]{A. Birsan$^{1,2}$}
\author[]{V. Kuncser$^{1}$}
\address{$^1$National Institute of Materials Physics, Atomistilor Str, No. 105 bis PO Box MG.7, 077125 Magurele, Romania. \\
 $^2$ University of Bucharest, Faculty of Physics, Atomistilor Str., No. 105 PO Box MG-11, 077125, Magurele, Romania}

\begin{abstract}
The half-metallic properties of a new and promising full-Heusler compound, Zr\raisebox{-.2ex}{\scriptsize 2}CoSn, are investigated by means of \textit{ab initio} calculations within the Density Functional Theory framework. The ferromagnetic ordered Hg\raisebox{-.2ex}{\scriptsize 2}CuTi-type crystal structure is energetically favorable and the optimized lattice parameter is 6.76$\dot{A}$. The total magnetic moment calculated is 3 $\mu_{B}/f.u.$ and follows a typical Slater-Pauling dependence. The half metallicity disappears if the unit cell volume is contracted by 5 $\%$.   
\end{abstract}


\maketitle

\section{Introduction}
\label{Introduction}
Advances in microelectronics and magnetic data storage devices, nowadays, depend on novel approaches of device fabrication based on the synergistic use of charge and spin dynamics of electrons in multifunctional materials. 
Various new device concepts have already found practical applications in magnetoelectronics or spintronics (e.q. read heads for magnetic recorders or nonvolatile memory components). For efficient spintronic devices, it is desirable to have nearly $100\%$ spin-polarized carrier injection. Since half metallic materials have electrons of only one spin state present, around Fermi level\cite{Groot1983}, they are promising candidates for use as spin injectors. One of the most interesting class of compounds with half metallic characteristics are Heusler alloys \cite{Heusler1903}, reported in literature in only two variants: the full-Heusler $X_{2}YZ$ compounds and half-Heusler $XYZ$ compounds; $X$ is a transition metal, $Y$ a transition metal or a rare-metal and $Z$ a main group element. The full-Heusler materials crystallize either in $Cu_{2}MnAl$ ($L2_{1}$) or in $Hg_{2}CuTi$-type structures. If $X$ atom is more electronegative than $Y$ is obtained the $Cu_{2}MnAl$ phase with $Fm\bar{3}m$ space group \cite{Kandpal2007JMMM}. The often called inverse Heusler structure ($Hg_{2}CuTi$-prototype) with $F\bar{4}3m$ space group  was reported when the $Y$ element more electronegative than $X$. The $X$ atoms in $Hg_{2}CuTi$-type structure, are placed in the Wyckoff positions 4a(0,0,0) and 4c(1/4,1/4,1/4) while $Y$ and $Z$ in 4b(1/2,1/2,1/2) and 4d(3/4,3/4,3/4), respectively \cite{Kandpal2007}. Even though no single set of properties can characterize the entire Heusler family, the magnetic behavior and multifunctional properties recently reported in literature, make these half-metallic systems to play an important role  in the research field of magnetic tunnel junctions \cite{vanEngen1983} and spintronics \cite{Tezuka2006,Sargolzaei2006}.  

Nowadays, promising materials which crystallize in $Hg_{2}CuTi$-prototype, e.g. $Mn_{2}$, $Ti_{2}$, $Sc_{2}$ - based Heusler compounds are intensively studied theoretically and experimentally, for  potential use in devices that can inject currents of high spin polarization \cite{Jabbar2013,Birsan2013JMMM,Zheng2012,Birsan2014Jallcom,Wei2014,Huang2012}.
In this paper, the electronic structure and magnetic properties calculated from first-principles investigations of a new proposed $Zr_{2}CoSn$ full-Heusler compound, are reported. Investigations based on Density Functional Theory (DFT) predict that the ideal $Zr_{2}CoSn$ system exhibits half metallic behavior and might be suitable for use in spintronics.

\section{Method of calculation}
\label{calculation}
The structural parameters of $Zr_{2}CoSn$ bulk material were determined using the Full-Potential Linearized Augmented Plane Wave (FPLAPW) method, as implemented in WIEN2K code \cite{Blaha}. 
The Perdew Burke Ernzerhof \cite{Perdew} Generalized Gradient Approximation (GGA) was employed for the  exchange-correlation parametrization. The muffin-tin radii $R_{MT}$ of 2.45 a.u., 2.52 a.u. and 2.38 a.u. were used for Zr, Co and Sn, respectively. 
Inside the muffin-tin spheres, lattice harmonics of up to $l=10$ were selected for the basis set. In the interstitial region, the plane wave cut-off value used was $K_{max}R_{MT} = 8$  (where $K_{max}$ is the maximum modulus for the reciprocal lattice vector). 
The tetrahedron method \cite{Blochl} with a grid containing 560 irreductible k points was selected in the irreducible part of the Brillouin zone (BZ) to construct the charge density in each self-consistency step. The cut off energy of -6 Ry defined the separation between the valence and core states. The charge convergence was checked versus the number of k points. The self-consistency was achieved when the total energy deviation was better than 0.01 mRy per cell and the integrated charge difference between two successive iterations less than $0.0001\:e/a.u.^{3} $.

\section{Results and Discussions}
\label{results}
\begin{figure}
 \begin{center}
    \includegraphics[scale=1]{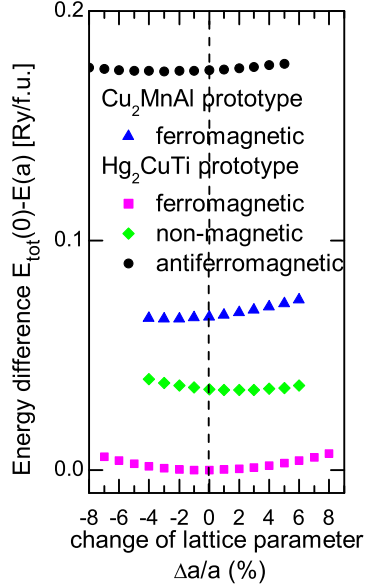}
  \end{center}
    \caption{Structural optimizations for $Zr_{2}CoSn$ compound, using the two structural prototypes:$Cu_{2}MnAl$ and $Hg_{2}CuTi$  }
        \label{fig:optimizareZr2CoSn}  
\end{figure}

Ab initio calculations were performed for bulk $Zr_{2}CoSn$ full-Heusler material, to investigate the existence and nature of magnetic properties in intercorrelation with the electronic structure. 
In general, experimental preparation and interpretation of true half-metallic compounds are still scarce, therefore from the beginning, structural optimization calculations needs to be considered to estimate the magnetic and structural stable phase by means of the total energy minimization. The ferromagnetic configurations of $Zr_{2}CoSn$ in the two crystal structure prototypes ($Cu_{2}MnAl$ and $Hg_{2}CuTi$), typical for the full-Heusler materials, and the antifferomagnetic and non-magnetic states of $Zr_{2}CoSn$ with $L2_{1}$ structure are the starting point of the calculations, as shown in Figure \ref{fig:optimizareZr2CoSn}. Based on the results, the ferromagnetic configuration of $Zr_{2}CoSn$, with $Hg_{2}CuTi$ structure is energetically favorable and has the lowest calculated total energy, therefore, the $Zr_{2}CoSn$ compound has to crystallize in the inverse Heusler structure with space group F-43m. Hence, the Wyckoff sequence considered for further calculations is 4a(Zr), 4c(Zr), 4b(Co) and 4d(Sn) and the calculated equilibrium lattice parameter was 6.76 $\dot{A}$ as corresponding to the $Zr_{2}CoSn$ ferromagnetic phase with $Hg_{2}CuTi$ -type structure.
\begin{figure}
 \begin{center}
         \includegraphics[scale=1]{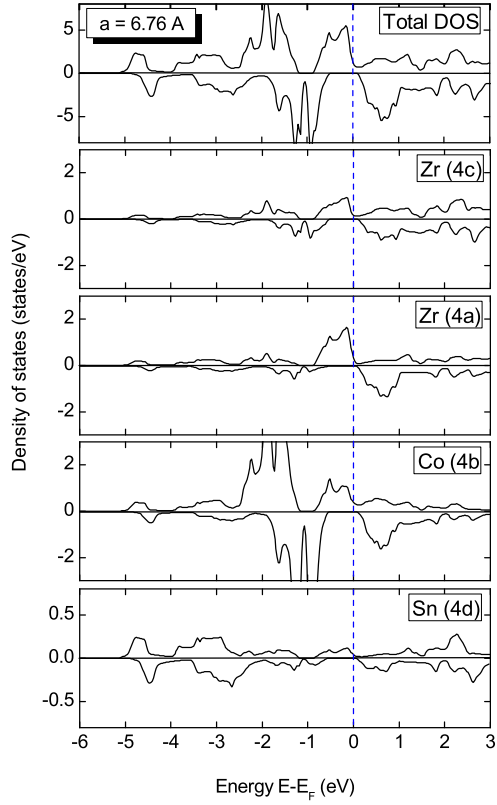}  
    \end{center}
    \caption{The spin projected total DOS and partial DOS calculated at predicted equilibrium lattice constant of $Zr_{2}CoSn$.}
       \label{fig:totaldosZr2CoSn}
\end{figure}
In the GGA scheme used in these investigations, half metallic properties are illustrated by the occurrence of an energy band gap in one of the spin channels and the integer total magnetic moment of the compound. The total and partial density of states of $Zr_{2}CoSn$ as function of energy difference $E_{tot}-E_{Fermi}$, performed at the equilibrium lattice constant are displayed in Figure \ref{fig:totaldosZr2CoSn}. The electronic structure reveals a metallic character in majority spin channel and a semiconducting behavior, with an energy gap, around Fermi level, in minority spin channel. It is concluded that $Zr_{2}CoSn$ exhibits half-metallic properties with a completely spin polarization in the ground state. 
\begin{figure}
 \begin{center}
    \includegraphics[scale=1.1]{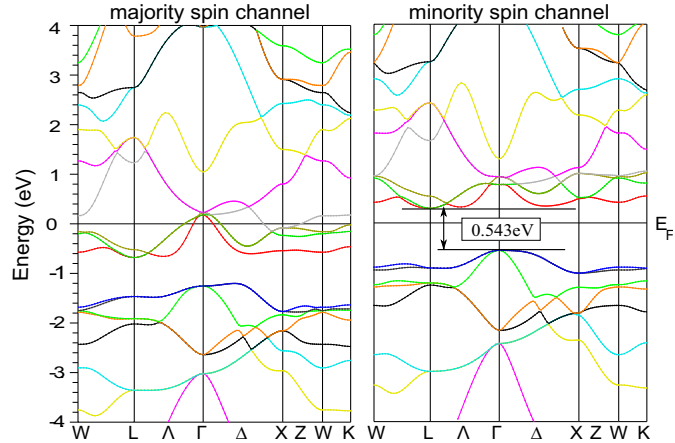} 
      \end{center}
   \caption{The band structure of $Zr_{2}CoSn$ for majority spin channel (spin-up) in the left panel and minority spin channel (spin-down) in the right panel for geometrical optimized structure}
    \label{fig:benziZr2CoSn}
\end{figure} 
In Figure \ref{fig:benziZr2CoSn} is illustrated the band structure of  $Zr_{2}CoSn$ full-Heusler alloy at optimized geometry. The density of states from majority spin channel is displayed in the left panel, while in the right panel of the figure is plotted the minority spin band structure where the complete absence of states at Fermi level implies the existence of an energy band gap of 0.543 eV. The indirect band gap is formed between the energy from the highest occupied states from valence band (VB)  at the $\Gamma$ point  and the lowest unoccupied states from conduction band (CB), at the $L$ point.  The spin flip, calculated as the gap from the highest valence band maximum of minority-spin to the Fermi level for $Zr_{2}CoSn$  compound at optimized lattice constant is 0.484 eV.  

\begin{figure}
 \begin{center}
    \includegraphics[scale=0.9]{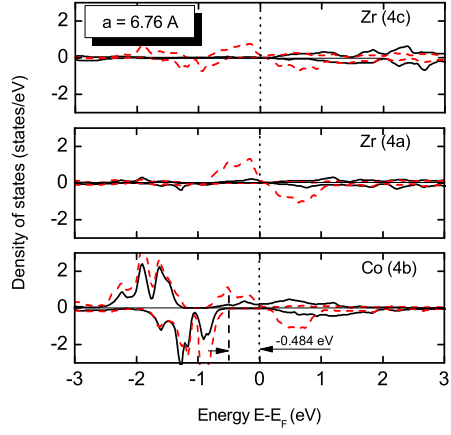}
      \end{center}
   \caption{The main partial densities of states at optimized lattice parameter of $Zr_{2}CoSn$, $d_{eg}$ and $d_{t2g}$ being indicated by solid and dashed line, respectively}
    \label{fig:sitedosZr2CoSn}
\end{figure}

The 3d electrons of Zr and Co atoms determine the energy band gap around the Fermi level. The density of states located above the Fermi level mainly comes from Zr(4a) atoms, while below the Fermi level the density of states of Co atoms has the main contribution(see Figure \ref{fig:sitedosZr2CoSn} ).   
\begin{figure}
 \begin{center}
    \includegraphics[scale=0.9]{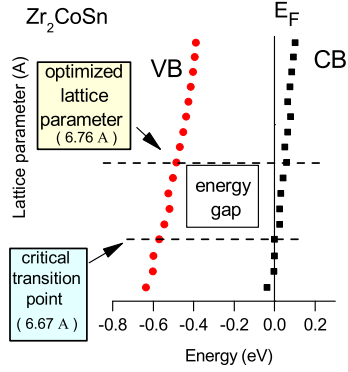}
      \end{center}
   \caption{The positions of the highest occupied states from valence band (solid squares) and of the lowest unoccupied states from the conduction band (solid circles) of total DOSs (minority spin channel) for $Zr_{2}CoSn$ as function of lattice parameter.}
    \label{fig:gapZr2CoSn}
\end{figure} 
The width of the energy gap and the position of Fermi level in minority spin band as function of lattice constant, are illustrated in  Figure \ref{fig:gapZr2CoSn}. The compound  is predicted to be an ideal candidate for spintronics, due to the existence of a gap in only one spin direction, for a large lattice parameter range. The half metallic properties disappears for a volume confinement of more than 5 $\%$ (e.g. corresponding to a lattice parameter of 6.67 $\dot{A}$). For further contraction, the Fermi Level falls within conduction band (CB), so that the compound becomes  a typical ferromagnet and the spin polarization decreases. However, a such critical volume confinement is high, making the compound a very stable one with respect to the polarization properties. 
\begin{figure}
 \begin{center}
    \includegraphics[scale=1]{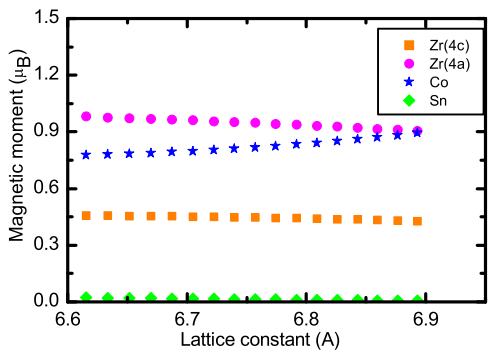}
      \end{center}
   \caption{The site-specific magnetic moments of $Zr_{2}CoSn$ compound vs lattice parameter.}
    \label{fig:sitemagneticmomentZr2CoSn}
\end{figure}
The calculated total magnetic moment of $Zr_{2}CoSn$ Heusler compound, with $Hg_{2}CuTi$ structure is 3 $\mu_{B}$ and follows the Slater-Pauling dependence $M_{t} = Z_{t}-18$ $\mu_{B}/f.u.$ \cite{Zheng2012} ($M_{t}$  is the total spin magnetic moments per formula unit cell and $Z_{t}$, the total number of valence electrons). The main contribution to the total magnetic moment comes from Zr(4a) and Co atoms, which are ferromagnetic coupled (see Figure \ref{fig:sitemagneticmomentZr2CoSn}). The spin-polarization calculations reveal that the site-resolved magnetic moments per atom, at the optimized lattice constant are 0.446,   0.946, 0.816 and  -0.013 $\mu_{B}$ for Zr(4c), Zr(4a), Co and Sn, respectively.  The magnetic moments of Zr (4a and 4c) atoms decrease, while the one of Co atoms increases, with increasing lattice constant.The different atomic environments determines the dissimilar local magnetic moments of zirconium atoms.  
\section{Conclusions}
\label{conclusions}
First principles investigations of electronic and  magnetic properties of $Zr_{2}CoSn$ full-Heusler alloy have been reported. The half-metallic behaviour in relation to the densities of states in the bulk material is reproduced for ground state. In the minority spin channel, the energy gap is 0.543 eV at the optimized lattice constant of 6.76$\dot{A}$. From applications point of view, the investigated material, $Zr_{2}CoSn$, is predicted to be suitable for spintronic devices due to high spin polarization. 
\section{Acknowledgments}
\label{acknowledgments}
A. Birsan would like to thank Dr. P. Palade for his support, and Dr. L. Ion for helpful discussions.
The authors acknowledge the financial support provided by the Romanian National Authority for Scientific Research through the CORE-PN45N projects. 
This work was also supported by the strategic grant POSDRU/159/1.5/S/137750, "Project Doctoral and Postdoctoral programs support for increased competitiveness in Exact Sciences research" cofinanced by the European Social Found within the Sectorial Operational Program Human Resources Development 2007-2013.
\bibliographystyle{elsarticle-harv}

\end{document}